\documentclass[11pt]{article}

\usepackage{times}
\usepackage{epic}
\usepackage{eepic}
\usepackage{amssymb}
\usepackage{amsmath}
\usepackage{amsbsy}
\usepackage{amsfonts}

\newtheorem{theo}{Theorem}

\newcommand{\qed}{\mbox{\rule{1.6mm}{4.3mm}}}

\newcommand{\one}{\mbox{\large{$1 \hspace{-0.95mm}  {\bf l}$}}}

\begin{document}

\title{Realization of a General Three-Qubit Quantum Gate}

\author{Farrokh Vatan and Colin P. Williams \\
      {\small Jet Propulsion Laboratory}  \\
      {\small California Institute of Technology} \\
      {\small 4800 Oak Grove Drive} \\
      {\small Pasadena, CA~~91109-8099}}
\date{ }

\maketitle

\begin{abstract}
We prove that a generic three-qubit quantum logic gate can be
implemented using at most 98 one-qubit rotations about the $y$- and $z$-axes
and 40 CNOT gates, beating an earlier bound
of 64 CNOT gates. 
\end{abstract}

\section{Introduction}
Currently, the quantum circuit model is the dominant paradigm for
describing the complexity of implementing a desired quantum
computation \cite{nielsen-chuang}. 
By decomposing an $n$-qubit quantum computation
into a sequence of one-qubit and two-qubit quantum gate operations,
one can characterize the complexity of a quantum computation via
the depth of the minimal quantum circuit that implements it.
Equivalently, a physicist can gain insight into the complexity of
performing certain desired manipulations on multi-partite states.
Clearly, any such statement of minimality must always be made with
respect to a particular universal family of quantum logic gates.
The most widely used family today is the set of all one-qubit gates
and CNOT gates \cite{barenco}. However, many equally good universal gate
families exist, and the choice of which family to use is
determined by whichever gates are the easiest to implement in a
chosen physical scheme for quantum computation. For example,
optics-based, superconductor-based, and spin-based quantum
computers would most likely use CNOT, $i$SWAP and $\sqrt{\mathrm{SWAP}}$
gates, respectively, as their preferred two-qubit entangling
operation. However, as there are simple relationships between one
entangling gate and another, insights into circuit complexity
based on the \{one-qubit, CNOT\} family are readily translated into the
other families. Consequently, in this paper we will concentrate on
characterizing the complexity for achieving arbitrary three-qubit
quantum computations using one-qubit gates and CNOT gates.
Furthermore, rather than allowing {\em any} one-qubit gate as a
primitive, we will only allow the use of phase gates and
rotations about the $y$- and $z$-axes.

It is known that a maximally general $n$-qubit quantum logic
operation can be implemented using $O(2^{2n})$ two-qubit gates
\cite{vartiainen}. Recently, this estimate has been made more precise for the
case of two-qubit quantum computations. Specifically, Vidal and
Dawson, Vatan and Williams and Shende, Markov and Bullock all
proved that a maximally general two-qubit gate can be achieved in a
quantum circuit that uses at most three CNOT gates. This result
remains valid if we replace CNOT with any other maximally
entangling two-qubit gate, such as $i$SWAP which is the more natural
entangling operation in superconductor-based quantum computers.
Moreover, if the available two-qubit operation, $U$, is less than
maximally entangling, than any two-qubit gate can be achieved with
at most six calls to $U$ \cite{vatan2}.

Unfortunately, the aforementioned bounds on the complexity of
achieving maximally general two-qubit quantum gates have not yet led
to similarly tight results for three-qubit gates. Three-qubit states are
especially interesting because they possess much richer
entanglement properties than two-qubit states
\cite{gingrich,kendon,carteretSudbery}. In particular, they ought
to allow physicists to gain a much deeper insight into the
distinction between nonlocality and entanglement
\cite{bennettQNLwE}. In order to investigate these states
experimentally it will be necessary to understand how to best
achieve arbitrary three-qubit gate operations. Such an understanding
might stimulate the development of new quantum information
processing protocols that rely upon tri-partite entangled states
\cite{yeo}.

To date, there have been relatively few results on the complexity
of implementing general three-qubit operations.
We refer to the most recent paper \cite{vartiainen}, where
it is shown, besides the other things, that a three-qubit operation can 
be implemented by using $136$ one-qubit gates and $64$ CNOT gates.
We improve this result by providing a computation that utilizes only
$98$ one-qubit gates and $40$ CNOT gates.

\section*{Notation}

We use the following simple notations for one-qubit gates:
\[ I=\one_2, \quad X=\sigma_x, \quad Y=\sigma_y, \quad Z=\sigma_z,
   \quad H=\mbox{$\frac{1}{\sqrt{2}}$} \begin{pmatrix} 1 & 1 \\ 1 & -1 \end{pmatrix}.   \]
For representing the tensor products of several one-qubit operations we apply the following
natural notation:   
\[ XXX = X \otimes X \otimes X, \quad YYX = Y \otimes Y \otimes X, \quad \mbox{etc.} \]
The following two three-qubit operations play the key rule in our construction:
\begin{align}
 N(a,b,c)   &= \exp\big( i(a\, XXZ+ b\, YYZ + c\, ZZZ )\big), \label{N-equ} \\
 M(a,b,c,d) &= \exp\big( i(a\, XXX+ b\, YYX + c\, ZZX + d\, IIX )\big). \label{M-equ}
\end{align}

\section{Construction}

The key element of our construction is the decomposition of a three-qubit unitary 
gate is the decomposition of such operations provided by Khaneja and Glaser
\cite{khaneja}. This is a general result which recursively reduces the computation 
of an arbitrary unitary operation on $n$ qubits, $U\in\mbox{\bf U}(2^n)$, into
of a sequence of operations on $n-1$ qubits and handful ``core'' operations on $n$ qubits.
For example, in the case of $n=2$ qubits, for $U\in\mbox{\bf U}(4)$ we have the
decomposition $U=K_1\,  A\, K_2$, where $K_1$ and $K_2$ belong to
$\mbox{\bf U}(2) \otimes \mbox{\bf U}(2)$, i.e., the space of operations on each
qubit separately, and $A$ is a two-qubit operation of the form 
$e^{a\, XX+b\, YY+ c\, ZZ}$. This decomposition is, in fact, the cornerstone of the
recent results on realization of two-qubit operations \cite{vatan, shende, vidal}.
Here we utilize the special form of this decomposition for the case of $n=3$ qubits.

\subsection{Khaneja-Glaser decomposition}

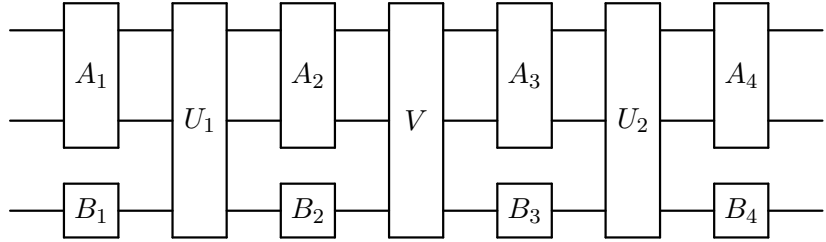
\begin{figure}[!ht]
\thicklines
\begin{center}
\unitlength=.24mm
\begin{picture}(450,110)(0,10)
\drawline(0,0)(30,0)
\drawline(0,50)(30,50)
\drawline(0,100)(30,100)
\drawline(30,-15)(30,15)(60,15)(60,-15)(30,-15)
\drawline(30,35)(30,115)(60,115)(60,35)(30,35)
\put(45,0){\makebox(0,0){$B_1$}}
\put(45,75){\makebox(0,0){$A_1$}}
\drawline(60,0)(90,0)
\drawline(60,50)(90,50)
\drawline(60,100)(90,100)
\drawline(90,-15)(90,115)(120,115)(120,-15)(90,-15)
\put(105,50){\makebox(0,0){$U_1$}}
\drawline(120,0)(150,0)
\drawline(120,50)(150,50)
\drawline(120,100)(150,100)
\drawline(150,-15)(150,15)(180,15)(180,-15)(150,-15)
\drawline(150,35)(150,115)(180,115)(180,35)(150,35)
\put(165,0){\makebox(0,0){$B_2$}}
\put(165,75){\makebox(0,0){$A_2$}}
\drawline(180,0)(210,0)
\drawline(180,50)(210,50)
\drawline(180,100)(210,100)
\drawline(210,-15)(210,115)(240,115)(240,-15)(210,-15)
\put(225,50){\makebox(0,0){$V$}}
\drawline(240,0)(270,0)
\drawline(240,50)(270,50)
\drawline(240,100)(270,100)
\drawline(270,-15)(270,15)(300,15)(300,-15)(270,-15)
\drawline(270,35)(270,115)(300,115)(300,35)(270,35)
\put(285,0){\makebox(0,0){$B_3$}}
\put(285,75){\makebox(0,0){$A_3$}}
\drawline(300,0)(330,0)
\drawline(300,50)(330,50)
\drawline(300,100)(330,100)
\drawline(330,-15)(330,115)(360,115)(360,-15)(330,-15)
\put(345,50){\makebox(0,0){$U_2$}}
\drawline(360,0)(390,0)
\drawline(360,50)(390,50)
\drawline(360,100)(390,100)
\drawline(390,-15)(390,15)(420,15)(420,-15)(390,-15)
\drawline(390,35)(390,115)(420,115)(420,35)(390,35)
\put(405,0){\makebox(0,0){$B_4$}}
\put(405,75){\makebox(0,0){$A_4$}}
\drawline(420,0)(450,0)
\drawline(420,50)(450,50)
\drawline(420,100)(450,100)

\end{picture} \end{center}
\caption{Khaneja-Glaser decomposition of a three-qubit unitary operation.}
\label{khaneja-fig}
\end{figure}

In \cite{khaneja} it is shown that every unitary operation on three qubits can be decomposed
as Figure~\ref{khaneja-fig},
where $A_j\in\mbox{\bf U}(4)$, $B_j\in\mbox{\bf U}(2)$,
$U_j=N(a_j,b_j,c_j)$ and $V=M(a,b,c,d)$ (see definitions (\ref{N-equ}) and (\ref{M-equ})).
We utilize the optimal construction of \cite{vatan}
for computing the two-qubit operations $A_j$'s. 
To complete our construction, we need computation of the three-qubit
operations $N(a,b,c)$ and $M(a,b,c,d)$. In the next sections we provide 
such computations.

\subsection{Computing $N(a,b,c)$}

Since the operations $XXZ$, $YYZ$, and $ZZZ$ are mutually commuting, we can write
\[ N(a,b,c) = \exp\big( i(a\, XXZ+ b\, YYZ )\big) \cdot
              \exp\big( i\, c\, ZZZ \big) . \]
Therefore, we break down the computation of $N(a,b,c)$ to computing unitary operations
\[ N_1(a,b)=\exp\big( i(a\, XXZ+ b\, YYZ )\big) \quad\mbox{and}\quad
   N_2(c)=\exp\big( i\, c\, ZZZ \big). \]
First, to compute $N_1(a,b)$, we introduce the following block-diagonal matrix
\begin{equation}
   P(a,b)=\begin{pmatrix}
     P_1 & & & \\  & P_1 & & \\ & & P_2 & \\ & & & P_2
     \end{pmatrix},
\label{Pab-equ}
\end{equation}
where
\[ P_1=\begin{pmatrix} \cos(a-b) & i\,\sin(a-b) \\
                       i\,\sin(a-b) & \cos(a-b) \end{pmatrix}
   \quad\mbox{and}\quad
   P_2=\begin{pmatrix} \cos(a+b) & i\,\sin(a+b) \\
                       i\,\sin(a+b) & \cos(a+b) \end{pmatrix}.\]
Then we have the decomposition of Figure~\ref{N1(a,b)-fig} for $N_1(a,b)$.
\begin{figure}[!ht]
\thicklines
\begin{center}
\unitlength=.24mm
\begin{picture}(570,110)(0,10)
\drawline(0,0)(60,0)
\drawline(0,50)(60,50)
\drawline(0,100)(270,100)
\put(30,50){\circle*{5}}
\put(30,100){\circle{16}}
\drawline(30,50)(30,108)
\drawline(60,0)(90,50)
\drawline(60,50)(90,0)
\drawline(90,0)(120,0)
\drawline(120,-15)(120,15)(150,15)(150,-15)(120,-15)
\put(135,0){\makebox(0,0){$H$}}
\drawline(150,0)(210,0)
\drawline(210,-15)(210,15)(240,15)(240,-15)(210,-15)
\put(225,0){\makebox(0,0){$H$}}
\drawline(240,0)(270,0)
\drawline(90,50)(270,50)
\put(180,50){\circle*{5}}
\put(180,0){\circle{16}}
\drawline(180,50)(180,-8)
\drawline(270,-15)(270,115)(300,115)(300,-15)(270,-15)
\put(285,50){\makebox(0,0){$P$}}
\drawline(300,0)(330,0)
\drawline(330,-15)(330,15)(360,15)(360,-15)(330,-15)
\put(345,0){\makebox(0,0){$H$}}
\drawline(360,0)(420,0)
\drawline(420,-15)(420,15)(450,15)(450,-15)(420,-15)
\put(435,0){\makebox(0,0){$H$}}
\drawline(450,0)(480,0)
\drawline(300,50)(480,50)
\put(390,50){\circle*{5}}
\put(390,0){\circle{16}}
\drawline(390,50)(390,-8)
\drawline(480,0)(510,50)
\drawline(480,50)(510,0)
\drawline(510,0)(570,0)
\drawline(510,50)(570,50)
\drawline(300,100)(570,100)
\put(540,50){\circle*{5}}
\put(540,100){\circle{16}}
\drawline(540,50)(540,108)

\end{picture} \end{center}
\caption{Decomposition of the unitary operation $N_1(a,b)$.}
\label{N1(a,b)-fig}
\end{figure}

Now to compute the operator $P(a,b)$, we use the technique of
Song-Williams \cite{colin} for decomposing block-diagonal unitary
matrices. The result is the circuit of Figure~\ref{P(a,b)-fig}.
\begin{figure}[!ht]
\thicklines
\begin{center}
\unitlength=.24mm
\begin{picture}(430,110)(0,10)
\drawline(0,0)(30,0)
\drawline(0,50)(430,50)
\drawline(0,100)(430,100)
\drawline(30,-15)(30,15)(90,15)(90,-15)(30,-15)
\put(60,0){\makebox(0,0){$R_z(-\frac{\pi}{2})$}}
\drawline(90,0)(120,0)
\drawline(120,-15)(120,15)(170,15)(170,-15)(120,-15)
\put(145,0){\makebox(0,0){$R_y(2a)$}}
\drawline(170,0)(230,0)
\drawline(230,-15)(230,15)(290,15)(290,-15)(230,-15)
\put(260,0){\makebox(0,0){$R_y(-2b)$}}
\drawline(290,0)(350,0)
\drawline(350,-15)(350,15)(400,15)(400,-15)(350,-15)
\put(375,0){\makebox(0,0){$R_z(\frac{\pi}{2})$}}
\drawline(400,0)(430,0)
\put(200,100){\circle*{5}}
\put(200,0){\circle{16}}
\drawline(200,100)(200,-8)
\put(320,100){\circle*{5}}
\put(320,0){\circle{16}}
\drawline(320,100)(320,-8)

\end{picture} \end{center}
\caption{Decomposition of the unitary operation $P(a,b)$.}
\label{P(a,b)-fig}
\end{figure}
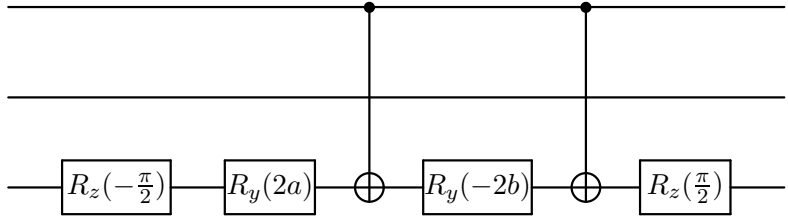

We combine the Figures~\ref{N1(a,b)-fig} and \ref{P(a,b)-fig} to obtain a decomposition
for $N_1(a,b)$. In this process, we commute the SWAP gate with the sequence
$\big( \one_2\otimes H\big)\cdot \mathrm{CNOT} \cdot \big( \one_2\otimes H\big)$, and
next we eliminate both SWAP gates by interchanging the rule of the second and third qubits.
The result is the circuit of Figure~\ref{N1(a,b)-2-fig}.
\begin{figure}[!ht]
\thicklines
\begin{center}
\unitlength=.24mm
\begin{picture}(560,110)(0,10)
\drawline(0,0)(60,0)
\drawline(0,50)(60,50)
\drawline(0,100)(560,100)
\put(30,50){\circle*{5}}
\put(30,100){\circle{16}}
\drawline(30,50)(30,108)
\drawline(60,35)(60,65)(120,65)(120,35)(60,35)
\put(90,50){\makebox(0,0){$R_z(-\frac{\pi}{2})$}}
\drawline(60,-15)(60,15)(90,15)(90,-15)(60,-15)
\put(75,0){\makebox(0,0){$H$}}
\drawline(90,0)(450,0)
\drawline(120,50)(180,50)
\put(150,50){\circle*{5}}
\put(150,0){\circle{16}}
\drawline(150,50)(150,-8)
\drawline(180,35)(180,65)(230,65)(230,35)(180,35)
\put(205,50){\makebox(0,0){$R_y(2a)$}}
\drawline(230,50)(290,50)
\put(260,100){\circle*{5}}
\put(260,50){\circle{16}}
\drawline(260,100)(260,42)
\drawline(290,35)(290,65)(350,65)(350,35)(290,35)
\put(320,50){\makebox(0,0){$R_y(-2b)$}}
\drawline(350,50)(450,50)
\put(380,100){\circle*{5}}
\put(380,50){\circle{16}}
\drawline(380,100)(380,42)
\put(420,50){\circle*{5}}
\put(420,0){\circle{16}}
\drawline(420,50)(420,-8)
\drawline(450,35)(450,65)(500,65)(500,35)(450,35)
\put(475,50){\makebox(0,0){$R_z(\frac{\pi}{2})$}}
\drawline(450,-15)(450,15)(480,15)(480,-15)(450,-15)
\put(465,0){\makebox(0,0){$H$}}
\drawline(480,0)(560,0)
\drawline(500,50)(560,50)
\put(530,50){\circle*{5}}
\put(530,100){\circle{16}}
\drawline(530,50)(530,108)

\end{picture} \end{center}
\caption{Decomposition of the unitary operation $N_1(a,b)$.}
\label{N1(a,b)-2-fig}
\end{figure}

Finally, the circuit of Figure~\ref{N2(c)-fig} computes the operator
$N_2(c)$.
\begin{figure}[!ht]
\thicklines
\begin{center}
\unitlength=.24mm
\begin{picture}(230,110)(0,10)
\drawline(0,0)(90,0)\drawline(140,0)(230,0)
\drawline(0,50)(230,50)
\drawline(0,100)(230,100)
\put(30,100){\circle*{5}}
\put(30,0){\circle{16}}
\drawline(30,100)(30,-8)
\put(200,100){\circle*{5}}
\put(200,0){\circle{16}}
\drawline(200,100)(200,-8)
\put(60,50){\circle*{5}}
\put(60,0){\circle{16}}
\drawline(60,50)(60,-8)
\put(170,50){\circle*{5}}
\put(170,0){\circle{16}}
\drawline(170,50)(170,-8)
\drawline(90,-15)(90,15)(140,15)(140,-15)(90,-15)
\put(115,0){\makebox(0,0){$R_z(2c)$}}

\end{picture} \end{center}
\caption{Decomposition of the unitary operation $N_2(c)$.}
\label{N2(c)-fig}
\end{figure}

Now by combining the circuits for $N_1(a,b)$ and $N_2(c)$ we obtain the circuit
of Figure~\ref{N(a,b,c)-fig} for computing the unitary operation $N(a,b,c)$.
\begin{figure}[!ht]
\thicklines
\begin{center}
\unitlength=.21mm
\begin{picture}(760,110)(0,10)
\drawline(0,0)(30,0)
\drawline(0,50)(30,50)
\drawline(0,100)(760,100)

\drawline(30,35)(30,65)(90,65)(90,35)(30,35)
\put(60,50){\makebox(0,0){\small $R_z(-\frac{\pi}{2})$}}
\drawline(30,-15)(30,15)(60,15)(60,-15)(30,-15)
\put(45,0){\makebox(0,0){\small $H$}}
\drawline(60,0)(450,0)
\drawline(90,50)(180,50)
\put(120,50){\circle*{5}}
\put(120,100){\circle{16}}
\drawline(120,50)(120,108)
\put(150,50){\circle*{5}}
\put(150,0){\circle{16}}
\drawline(150,50)(150,-8)
\drawline(180,35)(180,65)(230,65)(230,35)(180,35)
\put(205,50){\makebox(0,0){\small $R_y(2a)$}}
\drawline(230,50)(290,50)
\put(260,100){\circle*{5}}
\put(260,50){\circle{16}}
\drawline(260,100)(260,42)
\drawline(290,35)(290,65)(350,65)(350,35)(290,35)
\put(320,50){\makebox(0,0){\small $R_y(-2b)$}}
\drawline(350,50)(450,50)
\put(380,100){\circle*{5}}
\put(380,50){\circle{16}}
\drawline(380,100)(380,42)
\put(420,50){\circle*{5}}
\put(420,0){\circle{16}}
\drawline(420,50)(420,-8)
\drawline(450,35)(450,65)(500,65)(500,35)(450,35)
\put(475,50){\makebox(0,0){\small $R_z(\frac{\pi}{2})$}}
\drawline(450,-15)(450,15)(480,15)(480,-15)(450,-15)
\put(465,0){\makebox(0,0){\small $H$}}
\drawline(480,0)(620,0)
\drawline(500,50)(760,50)
\put(530,50){\circle*{5}}
\put(530,100){\circle{16}}
\drawline(530,50)(530,108)
\put(560,100){\circle*{5}}
\put(560,0){\circle{16}}
\drawline(560,100)(560,-8)
\put(730,100){\circle*{5}}
\put(730,0){\circle{16}}
\drawline(730,100)(730,-8)
\put(590,50){\circle*{5}}
\put(590,0){\circle{16}}
\drawline(590,50)(590,-8)
\put(700,50){\circle*{5}}
\put(700,0){\circle{16}}
\drawline(700,50)(700,-8)
\drawline(620,-15)(620,15)(670,15)(670,-15)(620,-15)
\put(645,0){\makebox(0,0){\small $R_z(2c)$}}
\drawline(670,0)(760,0)

\end{picture} \end{center}
\caption{Decomposition of the unitary operation $N(a,b,c)$.}
\label{N(a,b,c)-fig}
\end{figure}

\subsection{Computing $M(a,b,c,d)$}

Like the previous case, commutativity implies that
\[ M(a,b,c,d)= \exp\big( i(a\, XXX+ b\, YYX)\big) \cdot
               \exp\big( i(c\, ZZX)\big) \cdot \exp\big( i(d\, IIX)\big). \]
So we break down the decomposition of $M(a,b,c,d)$ to the task of computing the
following unitary operations:
\[ M_1(a,b)=\exp\big( i(a\, XXX+ b\, YYX)\big), \quad
   M_2(c)= \exp\big( i\,c\, ZZX\big), \quad \mbox{and}\quad
   M_3(d)=\exp\big( i\,d\, IIX\big). \]
At first step, we have decomposition of Figure~\ref{M1ab-fig} for the
unitary operation $M_1(a,b)$, where the operation $P$ is the same operation defined
by (\ref{Pab-equ}).
\begin{figure}[!ht]
\thicklines
\begin{center}
\unitlength=.24mm
\begin{picture}(360,110)(0,10)
\drawline(0,0)(180,0)
\drawline(0,50)(120,50)
\drawline(0,100)(120,100)
\put(30,0){\circle*{5}}
\put(30,100){\circle{16}}
\drawline(30,0)(30,108)
\put(60,100){\circle*{5}}
\put(60,50){\circle{16}}
\drawline(60,100)(60,42)
\put(90,0){\circle*{5}}
\put(90,50){\circle{16}}
\drawline(90,0)(90,58)
\drawline(120,50)(150,100)
\drawline(120,100)(150,50)
\drawline(150,50)(180,50)
\drawline(150,100)(180,100)
\drawline(180,-15)(180,115)(210,115)(210,-15)(180,-15)
\put(195,50){\makebox(0,0){$P$}}
\drawline(210,0)(360,0)
\drawline(210,50)(240,50)
\drawline(210,100)(240,100)
\drawline(240,50)(270,100)
\drawline(240,100)(270,50)
\drawline(270,50)(360,50)
\drawline(270,100)(360,100)
\put(300,0){\circle*{5}}
\put(300,100){\circle{16}}
\drawline(300,0)(300,108)
\put(330,100){\circle*{5}}
\put(330,50){\circle{16}}
\drawline(330,100)(330,42)

\end{picture} \end{center}
\caption{Decomposition of the unitary operation $M_1(a,b)$.}
\label{M1ab-fig}
\end{figure}

Then using the circuit of Figure~\ref{P(a,b)-fig} and eliminating the SWAP gates, we obtain
the circuit of Figure~\ref{M1ab-2-fig} for computing $M_1(a,b)$.
\begin{figure}[!ht]
\thicklines
\begin{center}
\unitlength=.24mm
\begin{picture}(590,110)(0,10)
\drawline(0,0)(120,0)
\drawline(0,50)(590,50)
\drawline(0,100)(590,100)
\put(30,0){\circle*{5}}
\put(30,100){\circle{16}}
\drawline(30,0)(30,108)
\put(60,100){\circle*{5}}
\put(60,50){\circle{16}}
\drawline(60,100)(60,42)
\put(90,0){\circle*{5}}
\put(90,50){\circle{16}}
\drawline(90,0)(90,58)
\drawline(120,-15)(120,15)(180,15)(180,-15)(120,-15)
\put(150,0){\makebox(0,0){$R_z(-\frac{\pi}{2})$}}
\drawline(180,0)(210,0)
\drawline(210,-15)(210,15)(260,15)(260,-15)(210,-15)
\put(235,0){\makebox(0,0){$R_y(2a)$}}
\drawline(260,0)(320,0)
\drawline(320,-15)(320,15)(380,15)(380,-15)(320,-15)
\put(350,0){\makebox(0,0){$R_y(-2b)$}}
\drawline(380,0)(440,0)
\drawline(440,-15)(440,15)(490,15)(490,-15)(440,-15)
\put(465,0){\makebox(0,0){$R_z(\frac{\pi}{2})$}}
\drawline(490,0)(590,0)
\put(290,50){\circle*{5}}
\put(290,0){\circle{16}}
\drawline(290,50)(290,-8)
\put(410,50){\circle*{5}}
\put(410,0){\circle{16}}
\drawline(410,50)(410,-8)
\put(520,0){\circle*{5}}
\put(520,100){\circle{16}}
\drawline(520,0)(520,108)
\put(560,100){\circle*{5}}
\put(560,50){\circle{16}}
\drawline(560,100)(560,42)

\end{picture} \end{center}
\caption{Decomposition of the unitary operation $M_1(a,b)$.}
\label{M1ab-2-fig}
\end{figure}

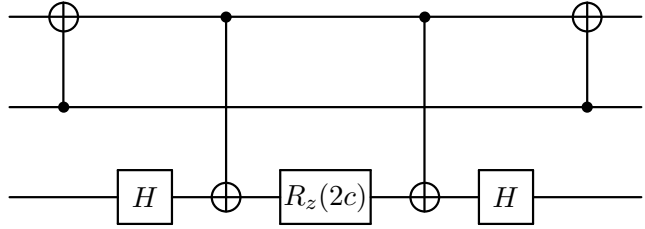
\begin{figure}[!ht]
\thicklines
\begin{center}
\unitlength=.24mm
\begin{picture}(350,110)(0,10)
\drawline(0,0)(60,0)
\drawline(0,50)(350,50)
\drawline(0,100)(350,100)
\put(30,50){\circle*{5}}
\put(30,100){\circle{16}}
\drawline(30,50)(30,108)
\drawline(60,-15)(60,15)(90,15)(90,-15)(60,-15)
\put(75,0){\makebox(0,0){$H$}}
\drawline(90,0)(150,0)
\put(120,100){\circle*{5}}
\put(120,0){\circle{16}}
\drawline(120,100)(120,-8)
\drawline(150,-15)(150,15)(200,15)(200,-15)(150,-15)
\put(175,0){\makebox(0,0){$R_z(2c)$}}
\drawline(200,0)(260,0)
\put(230,100){\circle*{5}}
\put(230,0){\circle{16}}
\drawline(230,100)(230,-8)
\drawline(260,-15)(260,15)(290,15)(290,-15)(260,-15)
\put(275,0){\makebox(0,0){$H$}}
\drawline(290,0)(350,0)
\put(320,50){\circle*{5}}
\put(320,100){\circle{16}}
\drawline(320,50)(320,108)

\end{picture} \end{center}
\caption{Decomposition of the unitary operation $M_2(c)$.}
\label{M2c-fig}
\end{figure}

The unitary operation $M_2(c)$ is simply decomposed as the circuit of
Figure~\ref{M2c-fig}.

\begin{figure}[!ht]
\thicklines
\begin{center}
\unitlength=.21mm
\begin{picture}(755,110)(0,10)
\drawline(0,0)(120,0)
\drawline(0,50)(755,50)
\drawline(0,100)(755,100)
\put(30,0){\circle*{5}}
\put(30,100){\circle{16}}
\drawline(30,0)(30,108)
\put(60,100){\circle*{5}}
\put(60,50){\circle{16}}
\drawline(60,100)(60,42)
\put(90,0){\circle*{5}}
\put(90,50){\circle{16}}
\drawline(90,0)(90,58)
\drawline(120,-15)(120,15)(150,15)(150,-15)(120,-15)
\put(135,0){\makebox(0,0){\small $S^*$}}
\drawline(150,0)(180,0)
\drawline(180,-15)(180,15)(210,15)(210,-15)(180,-15)
\put(195,0){\makebox(0,0){\small $A$}}
\drawline(210,0)(270,0)
\drawline(270,-15)(270,15)(300,15)(300,-15)(270,-15)
\put(285,0){\makebox(0,0){\small $B$}}
\drawline(300,0)(360,0)
\drawline(360,-15)(360,15)(390,15)(390,-15)(360,-15)
\put(375,0){\makebox(0,0){\small $S$}}
\drawline(390,0)(465,0)
\put(240,50){\circle*{5}}
\put(240,0){\circle{16}}
\drawline(240,50)(240,-8)
\put(330,50){\circle*{5}}
\put(330,0){\circle{16}}
\drawline(330,50)(330,-8)
\put(420,0){\circle*{5}}
\put(420,100){\circle{16}}
\drawline(420,0)(420,108)
\put(450,100){\circle*{5}}
\put(450,50){\circle{16}}
\drawline(450,100)(450,42)
\put(480,50){\circle*{5}}
\put(480,100){\circle{16}}
\drawline(480,50)(480,108)
\drawline(465,-15)(465,15)(495,15)(495,-15)(465,-15)
\put(480,0){\makebox(0,0){\small $H$}}
\drawline(495,0)(555,0)
\put(525,100){\circle*{5}}
\put(525,0){\circle{16}}
\drawline(525,100)(525,-8)
\drawline(555,-15)(555,15)(585,15)(585,-15)(555,-15)
\put(570,0){\makebox(0,0){\small $C$}}
\drawline(585,0)(645,0)
\put(615,100){\circle*{5}}
\put(615,0){\circle{16}}
\drawline(615,100)(615,-8)
\drawline(645,-15)(645,15)(675,15)(675,-15)(645,-15)
\put(660,0){\makebox(0,0){\small $D$}}
\drawline(675,0)(705,0)
\put(720,50){\circle*{5}}
\put(720,100){\circle{16}}
\drawline(720,50)(720,108)
\drawline(705,-15)(705,15)(735,15)(735,-15)(705,-15)
\put(720,0){\makebox(0,0){\small $H$}}
\drawline(735,0)(755,0)

\end{picture} \end{center}
\caption{Decomposition of the unitary operation $M(a,b,c,d)$, where
$A=R_y(2a)$, $B=R_y(-2b)$, $C=R_z(2c)$, $D=R_z(2d)$, and $S=R_z(\frac{\pi}{2})$.}
\label{Mabcd-fig}
\end{figure}
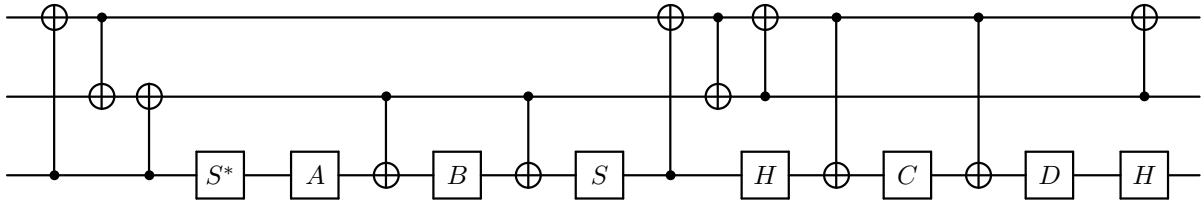

Finally, $M_3(d)=\one_4\otimes Q$, where
\[ Q=\begin{pmatrix} \cos(d) & i\, \sin(d) \\ i\, \sin(d) & \cos(d) \end{pmatrix}
    = R_z(\mbox{$\frac{\pi}{2}$}) \cdot R_y(2d) \cdot R_z(\mbox{-$\frac{\pi}{2}$}) . \]
Therefore, by putting together all these pieces, we find the circuit of Figure~\ref{Mabcd-fig}
for the unitary operation $M(a,b,c,d)$ (here we use the identity
$Q\cdot H= H \cdot R_z(2d)$).

\begin{theo}
Every unitary operation on three-qubits unitary can be computed by a circuit consisting of at 
most $98$ one-qubit phase gates and
rotations about the $y$- and $z$-axes and $40$ {\em CNOT} gates.
\end{theo}

{\bf Proof.}
To count the number of gates, we substitute the circuits of 
Figures~\ref{N(a,b,c)-fig} and \ref{Mabcd-fig} in 
Figure~\ref{khaneja-fig}. But before we start counting the number of gates right away, 
we should consider the possible ``cancellations''.
First note that the first (left) three gates of the circuit of
Figure~\ref{N(a,b,c)-fig} will be ``absorbed'' by their neighboring
gates $A_1$, $A_3$, $B_1$, and $B_3$.
Also the (only) $R_z(\frac{\pi}{2})$ gate of the same circuit commutes with
the gates on its right and will be ``absorbed'' by the gates $A_2$ and $A_4$.
Moreover, the last (right) three gates of the circuit of
Figure~\ref{Mabcd-fig} will be ``absorbed'' by their neighboring
$A_3$ and $B_3$ gates; and the left hand side gate $S^*$ commutes with the three
CNOT gates on its left and will be ``absorbed'' by the gate $B_2$.
Therefore, each of the operations $U_1$ and $U_2$ contributes $5$ one-qubit gates and
9 CNOT gates, and the operation $V$ contributes $6$ one-qubit gate and $10$ CNOT gates.

Finally, we utilize the identity $H=\sigma_z\cdot R_y(\frac{\pi}{2})$,
and the fact that every two-qubit unitary operation can be computed with at most
$15$ elementary one-qubit gates and $3$ CNOT gates \cite{vatan}. \qed

\subsection*{Acknowledgments}
The research described in this
paper was performed at the Jet Propulsion Laboratory (JPL),
California Institute of Technology, under contract with National
Aeronautics and Space Administration (NASA). We would like to
thank the sponsors, the National Security Agency (NSA), and the
Advanced Research and Development Activity (ARDA), for their
support.

\end{document}